\documentclass[preprint,onecolumn,nofootinbib]{revtex4}
\usepackage[colorlinks=true,linkcolor=blue,urlcolor=blue,filecolor=black,citecolor=red,pdfstartview=FitV,pdftitle={},pdfsubject={},pdfkeywords={},pdfpagemode=None,bookmarksopen=true]{hyperref}
\usepackage{graphicx}%Include figure files
\usepackage{amsmath}
\usepackage{booktabs} % 提供三线表所需的粗/细横线
\usepackage{amsfonts}
\usepackage{amssymb,ulem}
\usepackage{color,xcolor}%
\usepackage{CJK}
\usepackage{subfigure}
\usepackage{amsthm,amsmath,amssymb}
\usepackage{mathrsfs}
\usepackage{multirow} % 提供合并多行的功能

\usepackage{dcolumn}% Align table columns on decimal pointl
\usepackage{float}
\usepackage{bbm}
\usepackage{dsfont}
\newcommand{\red}{\textcolor[rgb]{1.00,0.00,0.00}}

\newcommand{\del}[1]{\red{\sout{#1}}}

\begin{document}
\title{Quasinormal Modes of a Black Hole in a Dehnen-Type Dark Matter Halo with Radial Pressure}

\author{Guoyang Fu$^{1}$}
\thanks{fuguoyang@yzu.edu.cn}
\author{Jiahao Zhou$^{1}$}
\thanks{jiahaozhou525@qq.com}
\author{Dan Zhang $^{1}$}
\thanks{danzhanglnk@163.com}
\author{Jian-Pin Wu$^{1}$}
\thanks{jianpinwu@yzu.edu.cn, corresponding author}
\affiliation{
$^1$\mbox{Center for Gravitation and Cosmology, College of Physical Science and Technology,} \mbox{Yangzhou University, Yangzhou 225009, China}\\
}

\begin{abstract}
We investigate the quasinormal modes (QNMs) of scalar and (axial) gravitational perturbations of a black hole spacetime surrounded by a Dehnen-type dark matter (DM) halo with nonvanishing radial pressure. Our results reveal that the real and imaginary parts of the QNMs exhibit distinct redshift coefficients with respect to the DM halo compactness $M/a_0$, in contrast to the vanishing radial pressure case. The magnitude of this effect is dominated by the profile parameter $\gamma$, and the damping rate is  more sensitive to the DM halo compactness than the oscillation frequency. In the eikonal limit, the redshift deviations are well captured by the light ring angular velocity $\Omega_c$ and the Lyapunov exponent $\lambda$.
\end{abstract}

\maketitle
%\tableofcontents

\section{Introduction}\label{sec-intro}

Black holes (BHs) are among the most profound predictions of general relativity (GR). Their existence has been observationally confirmed by gravitational-wave (GW) detections of binary black hole mergers \cite{LIGOScientific:2016aoc,LIGOScientific:2016lio,LIGOScientific:2016sjg} and further corroborated by observations of the shadows of M87* and Sgr A* made with the Event Horizon Telescope (EHT) \cite{EventHorizonTelescope:2022xqj,EventHorizonTelescope:2022wkp,EventHorizonTelescope:2019ths,EventHorizonTelescope:2019dse}. These observations open unprecedented opportunities for testing GR in the strong-field regime. Nonetheless, important open questions remain concerning quantum gravity and the nature of dark matter (DM) and dark energy. Exploring possible extensions of GR is therefore a natural avenue for addressing these fundamental questions.

In realistic astrophysical scenarios, BHs are not isolated vacuum systems but are embedded in complex matter distributions, with DM typically dominating the mass density on galactic scales. \cite{Gondolo:1999ef,Bertone:2005xv,Salucci:2018hqu,Bertone:2018krk,Barausse:2014tra,Cardoso:2019rou,Zwick:2022dih}. Because DM does not couple to electromagnetic radiation, its presence can only be probed through gravitational interactions \cite{Planck:2018vyg,Rubin:1970zza}. Therefore, establishing a theoretical framework that incorporates DM effects into BH spacetimes and identifying their observational signatures has become an important task in modern astrophysics. Recently, Cardoso \textit{et al.}~\cite{Cardoso:2021wlq} applied the Einstein cluster method to obtain a spherically symmetric BH solution embedded in a DM halo sourced by an anisotropic fluid with vanishing radial pressure. This solution is fully relativistic and self-consistently incorporates the gravitational backreaction of the DM halo on the central spacetime geometry. Related constructions with different DM profiles have since been developed~ \cite{AlBadawi:2024asn,Speeney:2024mas,Maeda:2024tsg,Shen:2023erj,Zhao:2026yis,Zare:2026lgt,Boltaev:2026prm,Konoplya:2022hbl,Ponglertsakul:2026upb,Fier:2026yjc,Shen:2024qbb}, including rotating generalizations~\cite{Fernandes:2025osu,Zhang:2024hjr,Fu:2026gcu,Yue:2026gju}. 

Quasinormal modes (QNMs) are the characteristic frequencies emitted by perturbed black holes, encoding the properties of the background spacetime and dominating the ringdown stage of GW emissions \cite{Konoplya:2011qq,Berti:2009kk,Kokkotas:1999bd}. 
Because QNMs are highly sensitive to the background geometry, any deviation from vacuum GR is expected to leave a distinct imprint on the QNM spectrum. QNMs have therefore been used extensively to test modified gravity \cite{Kobayashi:2012kh,Kobayashi:2014wsa,Ganguly:2017ort,Fu:2022cul,Antoniou:2024hlf,Antoniou:2024jku,Antoniou:2024gdf,Antoniou:2026vrm,Blazquez-Salcedo:2016enn,Molina:2010fb,Konoplya:2026cus,Xiong:2024urw,Xiong:2023usm} and quantum gravity \cite{Konoplya:2023hqb,Konoplya:2023ahd,Konoplya:2022hll,Konoplya:2022pbc,Fu:2023drp,Gong:2023ghh,Zhang:2024nny,Zhu:2024wic,Dong:2024ams,Zhang:2025ygb,Tang:2024txx}. More recently, the influence of matter environments around BHs on QNMs spectra has  attracted growing attention. In particular, relativistic corrections from the surrounding matter can deform the effective potential that governs the perturbation, leading to a redshift of the QNM spectrum \cite{Pezzella:2024tkf,Konoplya:2021ube,Feng:2025iao,Zhao:2026eti,Liu:2024xcd,Zhang:2021bdr}. This highlights the potential of using GW ringdown signals to constrain the density distribution and physical properties of DM.

A closely related development is the spectral instability of QNMs, which reveals that a small perturbation of the effective potential can lead to significantly changes in the QNM spectrum \cite{Nollert:1996rf,Nollert:1998ys}. Indeed, it has been shown that environmental effects, such as accretion disks, matter shells, or DM halos, can serve as natural sources of such perturbations, leading to a secondary bump in the effective potential that can significantly destabilize the QNM spectrum \cite{Jaramillo:2020tuu,Jaramillo:2021tmt,Cheung:2021bol,Berti:2022xfj,Courty:2023rxk,Cardoso:2024mrw,Spieksma:2024voy,Shen:2026hdl}. This suggests that environmental effects cannot be ignored in precision GW spectroscopy.

In this work, we focus on a special class of black hole solutions with a Dehnen-type DM halo density profile. We compute the QNMs of scalar and gravitational perturbations and analyze their dependence on the DM density.  The paper is organized as follows. In Sec. \ref{sec-II}, we briefly review the construction of black hole solutions with a Dehnen-type DM halo and derive the perturbation equations for scalar and gravitational fields. In Sec. \ref{sec-III}, we describe the numerical method used to compute the QNM frequencies, including the pseudo-spectral method and the WKB approximation. In Sec. \ref{sec-IV}, we present the QNM spectra and analyze the dependence of the redshift coefficients on the halo parameters. Finally, we summarize and discuss our results in Sec. \ref{sec-V}.

\section{Background spacetime and perturbation equations}\label{sec-II}

In this section, we first briefly review the construction of BH solutions with a Dehnen-type DM halo sourced by anisotropic matter. The density profile of galactic DM halo is given by the following general form \cite{Bosch_White_2010}: 
 \begin{eqnarray} \label{rho_DM}
\rho(r) = \rho_s \left( \frac{r}{a_0} \right)^{-\gamma} \left[ \left( \frac{r}{a_0} \right)^{\alpha} + 1 \right]^{\frac{\gamma - \beta}{\alpha}},
\end{eqnarray}
 %%%%ppp
where $\rho_s$ and $a_0$ are the scale density and the scale radius of DM halo, respectively. The parameters $(\alpha, \beta, \gamma)$ control the slope of the distribution, influencing the different spacetime regions. For the Dehnen density profile, we take $\alpha=1$ and $\beta=4$, with $0\leq \gamma < 3$. The corresponding static spherically symmetric metric can be written as  \cite{Boltaev:2026prm}:
\begin{eqnarray} \label{metric}
	ds^2=-A(r)dt^2+\frac{1}{B(r)}dr^2+r^2d\Omega^2\,,
\end{eqnarray}
%%%%ppp
where $A(r)$ and $B(r)$ are determined by solving the sourced Einstein equations
\begin{eqnarray} \label{Einstein_eq}
G_{\mu\nu}=8\pi T_{\mu\nu}\,.
\end{eqnarray}
%%%%ppp
The matter distribution is described by an anisotropic stress-energy tensor $T_{\mu\nu}$, given by
\begin{eqnarray} \label{Tdd}
T_{\mu\nu}=\left(\rho(r)+P_t(r) \right)u_\mu u_\nu + \left(P_r(r)-P_t(r)\right)k_\mu k_\nu+P_t(r) g_{\mu\nu}\,,
\end{eqnarray}
%%%%ppp
where $\rho(r)$ is the energy density, $P_r(r)$ and $P_t(r)$ correspond to the radial and tangential pressures. The vector $u^{\mu}$ is the timelike four-velocity of the fluid, while $k^\mu$ is a unit spacelike vector orthogonal to $u^{\mu}$, satisfying
\begin{eqnarray} \label{u_k}
u^{\mu} u_{\mu}=-1,\ k^{\mu}k_{\mu}=1,\ u^{\mu}k_{\mu}=0.
\end{eqnarray}
%%%%ppp

Imposing the anisotropic fluid condition $P_r(r)=-\rho(r)$, we can solve the Einstein field equations \eqref{Einstein_eq} and obtain the corresponding black hole solution as \cite{Boltaev:2026prm}:
\begin{eqnarray} \label{metric_2}
	ds^2=-f(r)dt^2+\frac{1}{f(r)}dr^2+r^2d\Omega^2\,,
\end{eqnarray}
%%%%ppp
with
\begin{eqnarray} \label{f_fun}
f(r)=1-\frac{2M_{BH}}{r}-\frac{4 M}{(r+a_0)(\gamma-3)}\left(\frac{r}{r+a_0}\right)^{2-\gamma}
\end{eqnarray}
%%%%ppp
where $M_{BH}$ is the mass of the central BH, and $M$ denotes the characteristic mass of the DM halo. It is worth noting that when the DM mass $M$ vanishes, the metric \eqref{f_fun} recovers the Schwarzschild black hole. 

To investigate the effects of the DM halo on the QNM spectrum, we consider scalar and gravitational perturbations. For the scalar field perturbation, the general covariant equation can be written as 
\begin{eqnarray} \label{scalar_eq}
\frac{1}{\sqrt{-g}}\partial_\mu \left( \sqrt{-g} g^{\mu\nu} \partial_\nu \Phi \right) = 0,
\end{eqnarray}
By substituting the metric \eqref{metric_2} into Eq. \eqref{scalar_eq} and separating the angular variables, the scalar perturbation equation can be reduced to the Schr\"odinger-like wave equation:
\begin{eqnarray} \label{Sch_eqs}
\frac{\partial^2 \Psi}{\partial r_*^2} + (\omega^2 - V_i)\Psi = 0,
\end{eqnarray}
where $r_*$ is the tortoise coordinate\del{,} defined as
\begin{equation}
dr_* =  \frac{dr}{f(r)}.
\end{equation}
The effective potential for the scalar field is given by
\begin{equation}
V_{s}(r) = f(r) \left[ \frac{\ell(\ell+1)}{r^2} + \frac{f'(r)}{r} \right].
\end{equation}

For the gravitational perturbations, we consider linear perturbations of both the metric and the stress-energy tensor. The perturbed field equations take the form
\begin{equation}
G^{(1)}_{\mu\nu} = 8\pi T^{(1)}_{\mu\nu}\,,
\end{equation}
where the superscript $(1)$ denotes the linear perturbation. In the Regge-Wheeler gauge \cite{Regge:1957td,Zerilli:1970se}, the perturbation equations can be separated according to parity into two independent sectors, corresponding to axial and polar perturbations. In this work, we focus on the axial perturbations and leave the polar sector for future study. For the axial sector, the matter perturbations can be eliminated, and the perturbation equation reduces to Eq. \eqref{Sch_eqs} with the effective potential 
\begin{equation}
V_{a}(r) = f(r) \left[ \frac{(\ell+2)(\ell-1)}{r^2} + \frac{2f(r)}{r^2}-\frac{rf'(r)}{r^2} \right].
\end{equation}

In the eikonal limit ($l\to \infty$), the centrifugal term dominates the effective potentials, and the scalar and axial gravitational potentials share the same leading order form
\begin{equation}
V_\infty(r)=f(r)\frac{\ell^2}{r^2}.
\end{equation}
The peak of this effective potential is located at the photon-sphere radius \(r_c\), which satisfies
\begin{equation}
r_c f'(r_c)-2f(r_c)=0,.
\end{equation}
Using the first order WKB approximation, the QNM frequencies can be expressed in terms of the parameters of the unstable circular null geodesic as
\begin{equation}
\omega_{ln}\simeq \Omega_c\, \ell - i\left(n+\frac12\right)|\lambda|,
\end{equation}
where the angular velocity $\Omega_c$ and Lyapunov exponent $\lambda$ are given by 
\begin{equation}
\Omega_c=\frac{\sqrt{f(r_c)}}{r_c},
\ \ \ 
\lambda=\sqrt{-\frac12 \frac{r_c^2}{f(r_c)}\left(\frac{d^2}{dr_*^2}\frac{f(r)}{r^2}\right)_{r=rc}}.
\end{equation}

\section{Numerical method}\label{sec-III}

BH QNMs carry rich information about the strong-field regime, since their complex frequencies are determined by the background geometry and the surrounding matter distribution. They are defined as the eigenfrequencies of the perturbation equation subject to purely ingoing waves at the event horizon and purely outgoing waves at spatial infinity. To solve this generalized eigenvalue problem, we employ the numerical pseudo-spectral method \cite{Boyd:Chebyshev,Jansen:2017oag}. For convenience, we first compactify the radial coordinate to a finite interval and render the perturbation equation linear in the frequency through the transformation
\begin{equation}
r=\frac{r_h}{u}, \ \ \Psi=e^{-i \omega r_*(u)}\psi .
\end{equation}
Here, $u=1$ and $u=0$ correspond to the event horizon and spatial infinity, respectively. Then, the perturbation equation \eqref{Sch_eqs} becomes
\begin{equation}\label{perturb_eq}
\psi''(u)+\left(\frac{2 i r_h \omega}{u^2 f(u)} + \frac{f'(u)}{f(u)}\right)\psi'(u)-\frac{1}{u}\left(\frac{2 \mathrm{i} r_h \omega}{u^2 f(u)}
+\frac{r_h^2 V_{\mathrm{eff}}(u)}{u^3 f(u)^2}
+\frac{f'(u)}{f(u)}\right)\psi(u)=0. 
\end{equation}
It is evident from Eq.~\eqref{perturb_eq} that only ingoing waves are present at the event horizon. We further introduce the following transformation to enforce the purely outgoing boundary condition at spatial infinity:
\begin{equation}\label{outgoing_BCs}
\psi(u)=e^{\frac{2i r_h \omega}{u}}u^{2i\omega r_h f'(0)} \delta\psi(u).
\end{equation}
Substituting the transformation \eqref{outgoing_BCs} into Eq.~\eqref{perturb_eq}, the final QNM equation can be rewritten in terms of $\delta\psi(u)$ as
\begin{equation}\label{QNM_eq}
\delta\psi''(u) + \lambda_0(u)\delta\psi'(u) + s_0(u)\delta\psi(u) = 0, 
\end{equation}
with
\begin{eqnarray}
\lambda_0(u) &=& \frac{f'(u)}{f(u)} + \frac{2ir_h \omega}{u}\left(2f'(0)+\frac{1}{uf(u)}-\frac{2}{u}\right), \nonumber \\
s_0(u) &=& \left(\frac{f'(u)}{uf(u)}+\frac{r_h^2 V_{eff}(u)}{u^4f(u)^2}\right)+\frac{2 i r_h \omega}{u}\left(\frac{f'(u)}{uf(u)}+\frac{1}{u^2f(u)}+\frac{f'(0)}{u}-\frac{f'(0)f'(u)}{f(u)}-\frac{2}{u^2}\right) \nonumber \\
&+&\frac{4r_h^2 \omega^2}{u^2} \left(f'(0)^2+\frac{f'(0)}{uf(u)}+\frac{1}{u^2}-\frac{1}{u^2f(u)}-\frac{2f'(0)}{u}\right).
\end{eqnarray}

With the QNM equation for $\delta\psi(u)$ in hand, we numerically solve it using the pseudo-spectral method. The key idea is to discretize the continuous variable on a set of collocation points, and to expand the unknown function $\delta\psi(u)$ in terms of the cardinal functions associated with the collocation points. Typically, we choose the Chebyshev grids and Lagrange cardinal functions
\begin{equation}
x_i = \cos\left(\frac{i}{N}\pi\right), \ \ C_j(x) = \prod_{\substack{i=0,i \neq j}}^{N} \frac{x-x_i}{x_j-x_i}, \quad i=0,\dots,N.
\end{equation}
The expansion in terms of these cardinal functions enables the construction of the differentiation matrices $D_{ij}^{(n)}=C_j^{(n)}(x_i)$. This reduces the continuous differential problem to a finite-dimensional algebraic system, which for the QNM equation takes the form of a quadratic eigenvalue problem in $\omega$. 

Further, the QNM equation \eqref{QNM_eq} can be generalized to a general polynomial dependence on frequency 
\begin{equation}\label{PS_matrix}
\mathcal{M}(\omega)\delta\psi=(\tilde{M}_0+\omega \tilde{M}_1+\omega^2 M_2+...+\omega^p \tilde{M}_p)\delta\psi=0.
\end{equation}
For a nontrivial solution, the determinant of $\mathcal{M}(\omega)$ must vanish,
\begin{equation}
\text{det}\mathcal{M}(\omega)=0.
\end{equation}
Solving this secular equation yields the eigenvalues $\omega$, which correspond to the QNM frequencies, while the associated eigenvectors describe the mode profiles. However, this method is computationally inefficient because it requires sweeping the complex $\omega$-plane and repeatedly evaluating det$\mathcal{M}(\omega)$ over a grid of frequency points to locate its roots. 

An efficient alternative is to linearize the polynomial eigenvalue problem into a generalized eigenvalue problem. Introducing the auxiliary vector $\mathcal{Q}=(\delta\psi, \omega \delta\psi,...,\omega^{p-1}\delta\psi)^\text{T}$, Eq.~\eqref{PS_matrix} can be recast as
\begin{equation}
(M_0+\omega M_1)\mathcal{Q}=0
\end{equation}
where $M_0$ and $M_1$ are the block companion matrices defined as
\begin{equation*}
M_0 = \begin{pmatrix}
\tilde{M}_0 & \tilde{M}_1 & \tilde{M}_2 & \cdots & \tilde{M}_{p-1} \\
0 & \mathbf{1} & 0 & \cdots & 0 \\
0 & 0 & \mathbf{1} & \cdots & 0 \\
\vdots & \vdots & \vdots & \ddots & \vdots \\
0 & 0 & 0 & 0 & \mathbf{1}
\end{pmatrix},
\quad
M_1 = \begin{pmatrix}
0 & 0 & 0 & \cdots & 0 & \tilde{M}_p \\
-\mathbf{1} & 0 & 0 & \cdots & 0 & 0 \\
0 & -\mathbf{1} & 0 & \cdots & 0 & 0 \\
\vdots & \vdots & \vdots & \ddots & \vdots & \vdots \\
0 & 0 & 0 & \cdots & -\mathbf{1} & 0
\end{pmatrix}.
\end{equation*}
This generalized eigenvalue problem can be solved directly with a standard eigenvalue solver, thereby improving the computational efficiency \cite{Boyd:Chebyshev,Jansen:2017oag}.

To further validate our results, we also independently compute the QNM frequencies using the WKB approximation. This method expands the effective potential around its maximum and, by matching the WKB solutions across the two turning points, directly relates the QNM frequencies to the effective potential and its derivatives at the maximum \cite{1985ApJ291L33S,Iyer:1986np,Guinn:1989bn,Konoplya:2004ip}. In recent years, the WKB method has been extended to higher orders and adopted Padé approximants, leading to significant improvements in both computational accuracy and efficiency \cite{Matyjasek:2017psv,Konoplya:2019hlu,Matyjasek:2026yiu}. In this work, we adopt the higher-order WKB approximation with Padé resummation of Ref. \cite{Konoplya:2019hlu} and require agreement with the pseudo-spectral results at the percent level.

\section{QUASINORMAL MODES}\label{sec-IV}

\begin{figure}[H]
	\centering
	\includegraphics[width=0.45\textwidth]{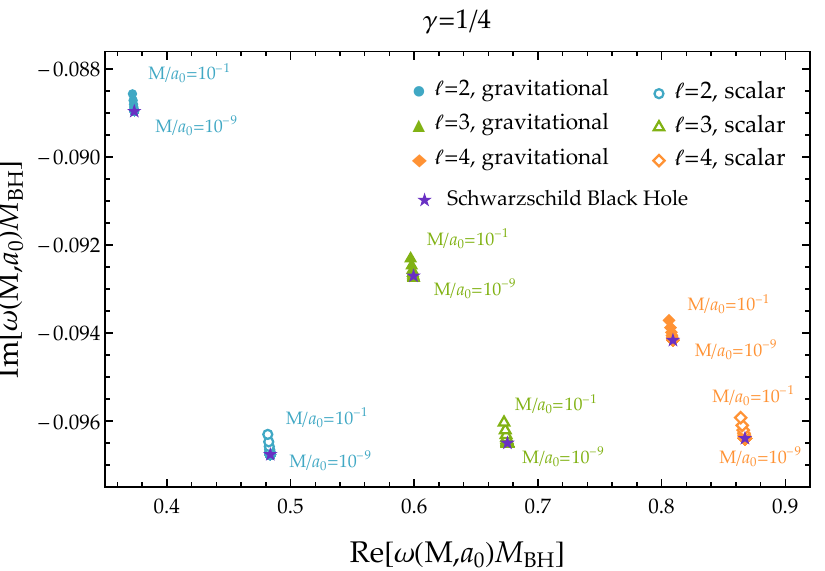} \hspace{0.2mm}	
    \includegraphics[width=0.45\textwidth]{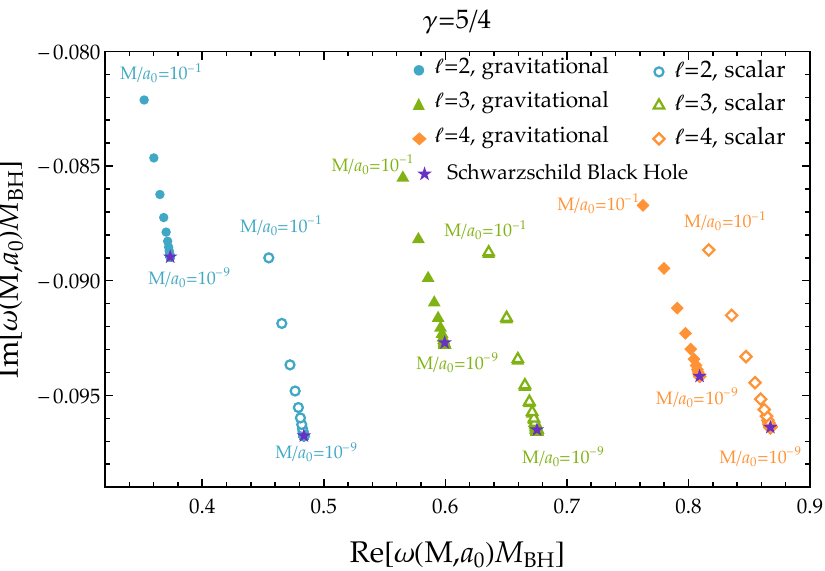}
	\caption{QNM spectra for gravitational and scalar perturbations as functions of the halo compactness $M/a_0$ for different $\ell$ and $\gamma$. }
	\label{FIG1}
\end{figure}

\begin{figure}[H]
	\centering
	\includegraphics[width=0.36\textwidth]{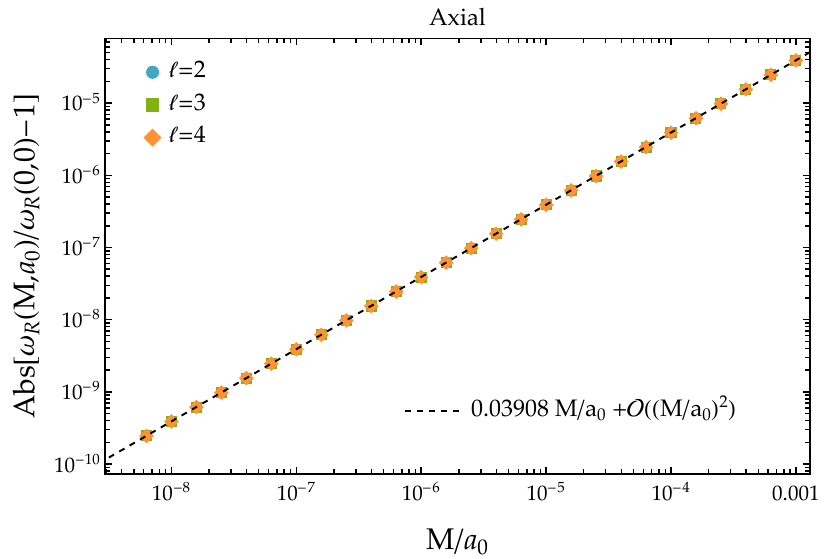} \hspace{0.2mm}	
    \includegraphics[width=0.36\textwidth]{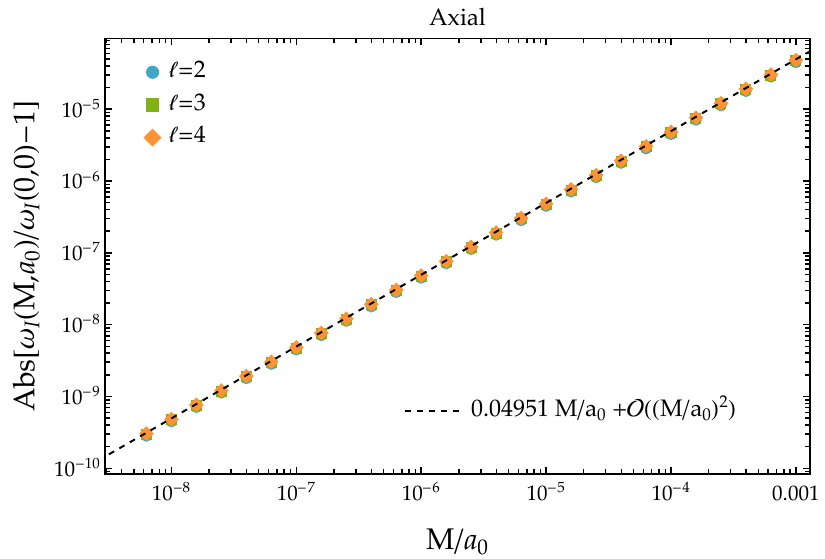}
    \includegraphics[width=0.36\textwidth]{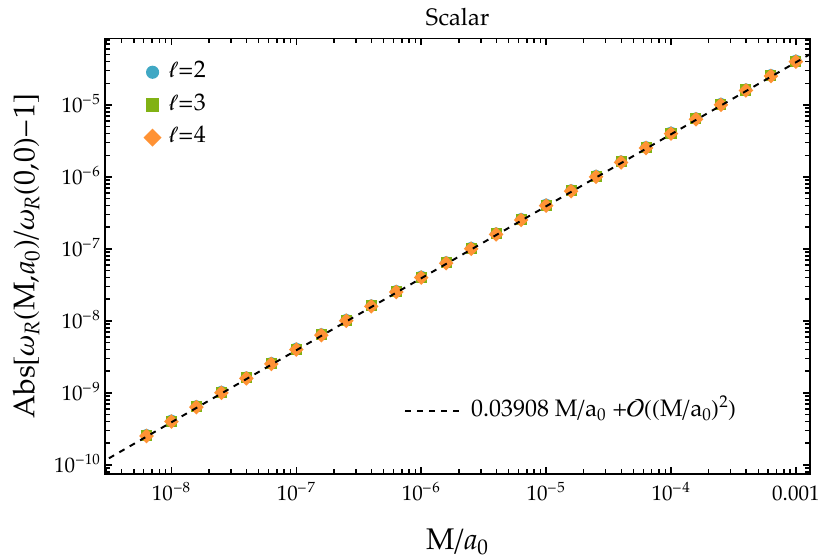} \hspace{0.2mm}	
    \includegraphics[width=0.36\textwidth]{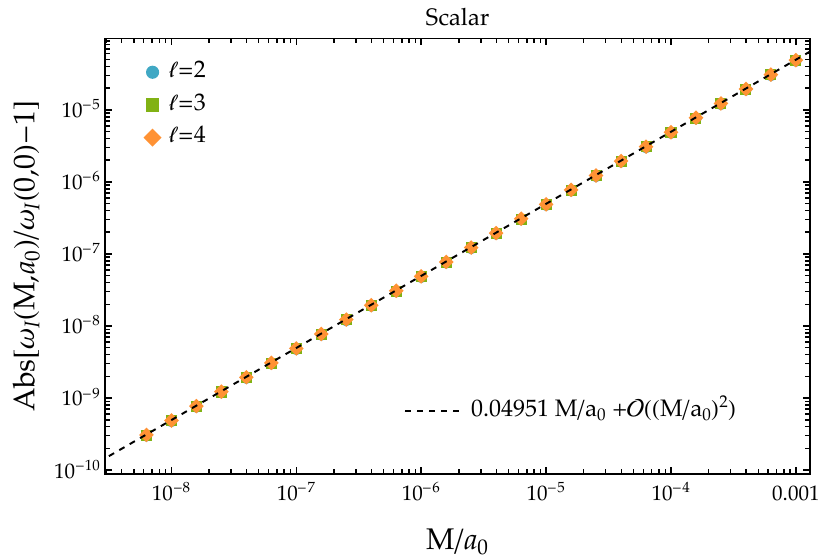}
	\caption{Relative deviations of the real and imaginary parts of the QNMs for the DM model with $\gamma=1/4$ compared to the Schwarzschild case.  The black dashed line correspond to the numerical fit to the relative deviation in the eikonal limit. The real part is given in terms of the angular velocity at the light ring as $\omega_R\simeq \ell \Omega_c$, while the imaginary part is given in terms of the Lyapunov exponent as $\omega_I\simeq -\left(n+\frac{1}{2}\right)|\lambda_c|$. }
	\label{FIG2}
\end{figure}

\begin{figure}[H]
	\centering
	\includegraphics[width=0.36\textwidth]{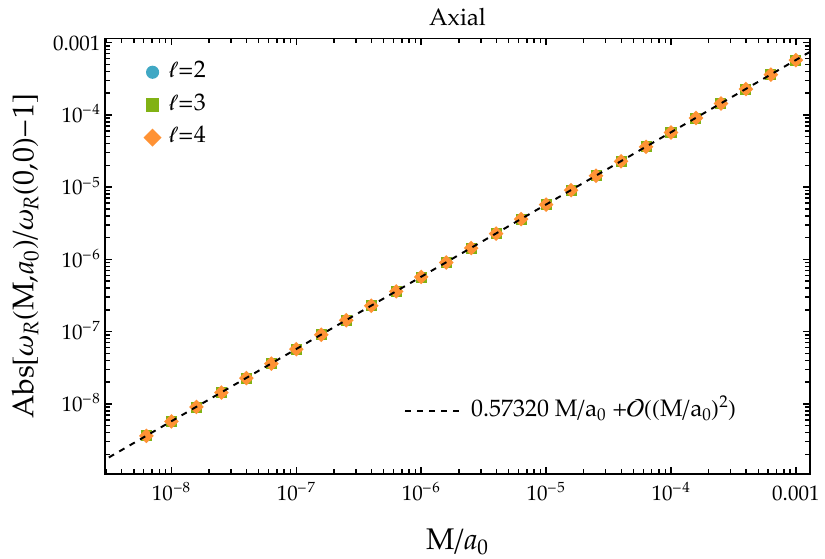} \hspace{0.2mm}	
    \includegraphics[width=0.36\textwidth]{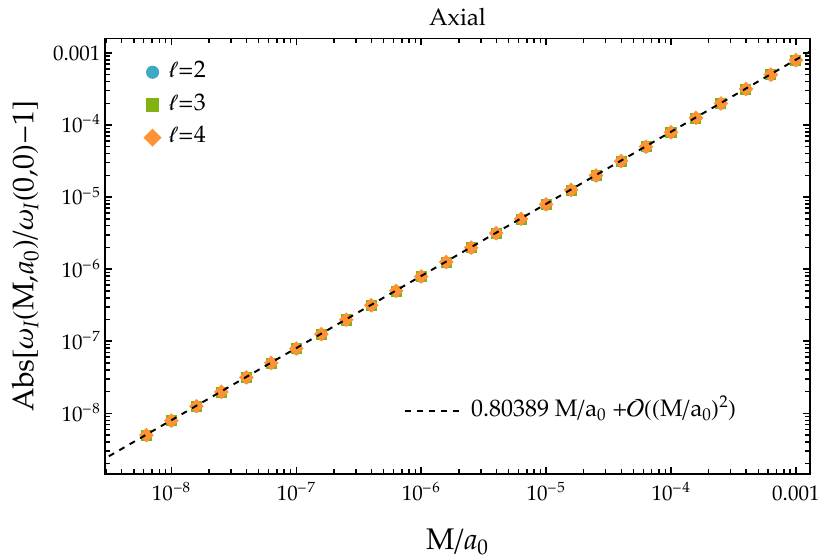}
    \includegraphics[width=0.36\textwidth]{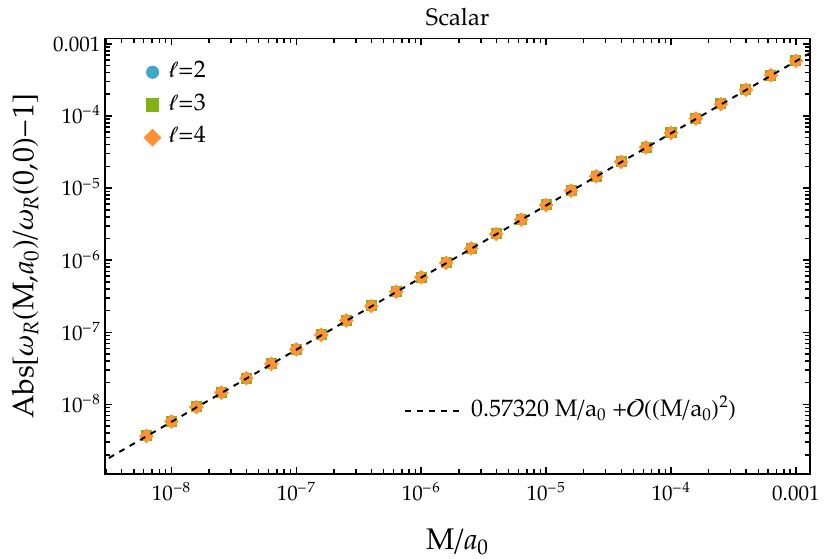} \hspace{0.2mm}	
    \includegraphics[width=0.36\textwidth]{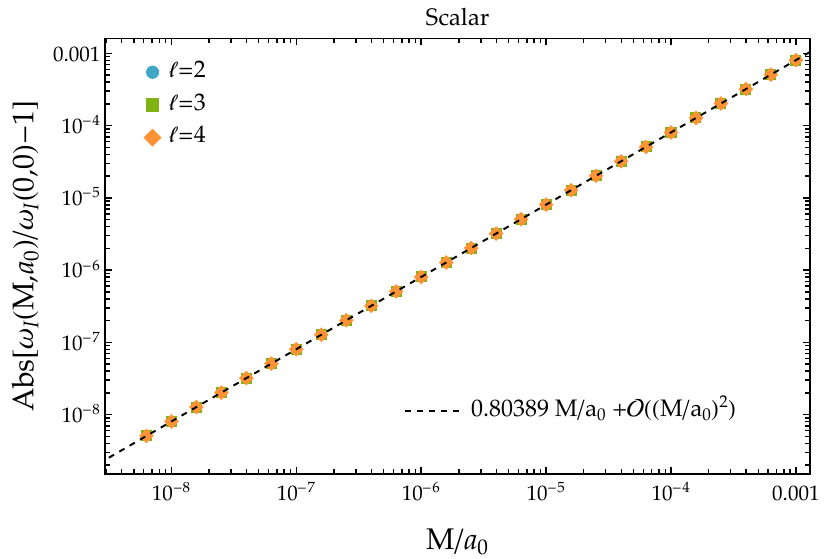}
	\caption{Relative deviations of the real and imaginary parts of the QNMs for the DM model with $\gamma=5/4$ compared to the Schwarzschild case.  The black dashed line correspond to the numerical fit to the relative deviation in the eikonal limit. The real part is given in terms of the angular velocity at the light ring as $\omega_R\simeq \ell \Omega_c$, while the imaginary part is given in terms of the Lyapunov exponent as $\omega_I\simeq -\left(n+\frac{1}{2}\right)|\lambda_c|$. }
	\label{FIG3}
\end{figure}

We now present the QNM spectra of scalar and gravitational perturbations in the presence of a Dehnen-type DM halo. Figures \ref{FIG1}–\ref{FIG3} summarize the main numerical results for the QNM frequencies. Figure~\ref{FIG1} shows the QNM frequencies in the complex plane for different values of the halo compactness $M/a_0$ and angular number $\ell$, with the left and right panels corresponding to $\gamma=1/4$ and $\gamma=5/4$, respectively. The imaginary parts are always negative, which indicates that the spacetime is stable against the linear perturbations considered here. As the halo compactness $M/a_0$ increases, both the real part and the absolute value of the imaginary part decrease. This trend is independent of the angular number $\ell$ and the type of perturbation. For fixed $M/a_0$ and $\ell$, scalar modes have larger real parts and larger $|\mathrm{Im}\,\omega|$ than axial gravitational modes, indicating higher oscillation frequencies and stronger damping. In addition, we observed that the QNMs exhibit more pronounced deviations from the Schwarzschild case for larger $\gamma$.

%%%%pppppp
%%%%%p

Although the scalar and axial frequencies differ in magnitude, Fig.~\ref{FIG1} shows that they share the same qualitative dependence on $M/a_0$. To quantify the effect of the DM halo, we analyze the relative deviations of the real and imaginary parts of the QNM frequencies from their Schwarzschild values, as shown in Figs.~\ref{FIG2} and~\ref{FIG3}. We find that the redshifts of the QNM spectra approximately follow the same relation, 
\begin{equation}\label{redshift}
\frac{\omega_{\text{R/I}}(M,a_0)}{\omega_{\text{R/I}}(0,0)}\simeq1 - \mathcal{C} \frac{M}{a_0}+\mathcal{O}\left(\frac{M}{a_0}\right)^2,
\end{equation}
where the redshift coefficient $\mathcal{C}$ depends on the DM halo parameters $\gamma$ and the DM profiles considered. In Ref.~\cite{Pezzella:2024tkf}, an anisotropic DM distribution with vanishing radial pressure yields equal redshift coefficients for the real and imaginary parts. However, in the present case, nonzero radial pressure breaks the equality between the redshift coefficients for the real and imaginary parts, because the radial pressure modifies the local geometry near the light ring so that $\Omega_c$ and $\lambda_c$ no longer share the same redshift factor. In all cases considered here, the redshift coefficient of the imaginary part exceeds that of the real part, showing that the damping rate is more sensitive to the DM halo than the oscillation frequency. Nevertheless, in the eikonal limit, the shifts remain well described by the light-ring quantities. Specifically, the fitted redshift coefficients are listed in Table~\ref{tab:redshift_coeff}.

%%%%pppppp
%%%%%p

%%%%pppppp
%%%%%p

\begin{table}[htbp]
    \centering
    \caption{Redshift coefficients \(\mathcal{C}\) obtained by numerical fitting with Eq.~\eqref{redshift} for different perturbation types, angular numbers \(\ell\), and DM halo parameters \(\gamma\).}
    \label{tab:redshift_coeff}
    \begin{tabular}{c|c|c|ccc}
        \hline 
        \(\gamma\) & \textbf{Perturbation} &  & \(\ell=2\) & \(\ell=3\) & \(\ell=4\) \\
        \hline 
        \multirow{4}{*}{\(1/4\)} 
          & \multirow{2}{*}{Axial}  & Re & 0.03929 & 0.03892 & 0.03894 \\
          &                         & Im & 0.04466 & 0.04728 & 0.04824 \\
        \cline{2-6} 
          & \multirow{2}{*}{Scalar} & Re & 0.04188 & 0.04053 & 0.03996 \\
          &                         & Im & 0.04719 & 0.04832 & 0.04879 \\
        \hline 
        \multirow{4}{*}{\(5/4\)} 
          & \multirow{2}{*}{Axial}  & Re & 0.56829 & 0.56827 & 0.56988 \\
          &                         & Im & 0.76904 & 0.78437 & 0.79221 \\
        \cline{2-6}
          & \multirow{2}{*}{Scalar} & Re & 0.59302 & 0.58337 & 0.57939 \\
          &                         & Im & 0.80303 & 0.80350 & 0.80366 \\
        \hline 
    \end{tabular}
\end{table}

\section{discussion}\label{sec-V}

%The QNM spectrum in vacuum is determined by the spacetime geometry, but the presence of a matter environment inevitably leaves its imprint on it. In this work, we consider a special class of analytic Schwarzschild-like black hole solutions that describe a BH surrounded by a Dehnen-type DM halo with nonvanishing radial pressure. We then study both the scalar and axial gravitational perturbations. By separating variables, we further reduce the equations to Schrödinger-like forms and give the corresponding effective potentials. To compute the QNMs, we employ the numerically pseudo-spectral method and verify the results with the WKB approximation. We find that the DM environment around the BH leads to a systematic redshift of the QNM spectrum, which is independent of the perturbation type and depends only on the system parameters, such as the halo compactness $M/a_0$ and profile parameter $\gamma$. 

%%%%% ===== New Discussion (marked in red) =====
The QNM spectrum of an isolated black hole is determined solely by the spacetime geometry. In realistic astrophysical environments, however, surrounding matter modifies the effective potential governing linear perturbations and thereby imprints a characteristic shift on the ringdown frequencies. In this work, we have investigated an analytic Schwarzschild-like BH immersed in a Dehnen-type DM halo with nonvanishing radial pressure, $P_r=-\rho$, focusing on massless scalar and axial gravitational perturbations. After casting the master equations to Schr\"odinger-like form, we computed the QNMs using a pseudo-spectral method and cross-checked the results with the higher-order WKB approximation supplemented by Pad\'e resummation.

Our numerical results show that the imaginary parts of the QNM frequencies remain negative throughout the parameter space explored here, indicating linear stability with respect to the perturbations under consideration. As the halo compactness $M/a_0$ increases, both $\mathrm{Re}\,\omega$ and $|\mathrm{Im}\,\omega|$ decrease, corresponding to a systematic redshift of the oscillation frequency and weaker damping. This trend is independent of the multipole number $\ell$ and is qualitatively the same for scalar and axial modes, although at fixed parameters scalar modes typically exhibit larger $\mathrm{Re}\,\omega$ and larger $|\mathrm{Im}\,\omega|$ than axial gravitational modes. The magnitude of the environmental correction is controlled primarily by the profile parameter $\gamma$: the deviations from the Schwarzschild spectrum are substantially more pronounced for $\gamma=5/4$ than for $\gamma=1/4$, as quantified by the fitted redshift coefficients in Table~\ref{tab:redshift_coeff}.

A central finding of this analysis concerns the redshift coefficients themselves. By fitting the relative deviations of $\omega_{\mathrm{R}}$ and $\omega_{\mathrm{I}}$ to Eq.~\eqref{redshift}, we find that the real and imaginary parts of the QNM frequency acquire unequal redshift coefficients $\mathcal{C}$. This stands in contrast to anisotropic DM models with vanishing radial pressure, for which the two coefficients coincide~\cite{Pezzella:2024tkf}. The nonzero radial pressure modifies the local geometry near the light ring, so that the orbital frequency $\Omega_c$ and the Lyapunov exponent $\lambda$ no longer share a common redshift factor. In all cases considered here, the redshift coefficient of the imaginary part exceeds that of the real part, implying that the damping rate is more sensitive to the DM halo than the oscillation frequency. Nonetheless, in the eikonal limit the shifts remain well captured by the light-ring quantities $\Omega_c$ and $\lambda$, reinforcing the geometric interpretation of the leading environmental corrections.

Finally, it is worth noting that we have restricted our analysis to axial gravitational perturbations and have left the polar sector for future work. Polar modes couple more directly to matter perturbations and may therefore exhibit additional environmental signatures. Extending the analysis to rotating backgrounds would further clarify the observational implications for GW spectroscopy of astrophysical black holes. Finally, environmental deformations of the effective potential are known to be closely connected with the spectral instability of QNMs. Exploring how a Dehnen-type halo with radial pressure affects overtones and late-time signals remains an interesting avenue for future work.

\acknowledgments

This work is supported by the Natural Science Foundation of China under Grant Nos. 12505078, 12505085, 12375055, the China Postdoctoral Science Foundation (Grant No. 2025T180931), and the Jiangsu Funding Program for Excellent Postdoctoral Talent (Grant No. 2025ZB705).

%%%%%%%%%%%%%%%%%%%  检查下国自然基金号  多了一个  我的是078  %%%%%%%%%%%%%%%%%%%

\bibliographystyle{style1}
\bibliography{Ref}

@article{LIGOScientific:2016aoc,
    author = "Abbott, B. P. and others",
    collaboration = "LIGO Scientific, Virgo",
    title = "{Observation of Gravitational Waves from a Binary Black Hole Merger}",
    eprint = "1602.03837",
    archivePrefix = "arXiv",
    primaryClass = "gr-qc",
    reportNumber = "LIGO-P150914",
    doi = "10.1103/PhysRevLett.116.061102",
    journal = "Phys. Rev. Lett.",
    volume = "116",
    number = "6",
    pages = "061102",
    year = "2016"
}

@article{LIGOScientific:2016lio,
    author = "Abbott, B. P. and others",
    collaboration = "LIGO Scientific, Virgo",
    title = "{Tests of general relativity with GW150914}",
    eprint = "1602.03841",
    archivePrefix = "arXiv",
    primaryClass = "gr-qc",
    reportNumber = "LIGO-P1500213",
    doi = "10.1103/PhysRevLett.116.221101",
    journal = "Phys. Rev. Lett.",
    volume = "116",
    number = "22",
    pages = "221101",
    year = "2016",
    note = "[Erratum: Phys.Rev.Lett. 121, 129902 (2018)]"
}

@article{LIGOScientific:2016sjg,
    author = "Abbott, B. P. and others",
    collaboration = "LIGO Scientific, Virgo",
    title = "{GW151226: Observation of Gravitational Waves from a 22-Solar-Mass Binary Black Hole Coalescence}",
    eprint = "1606.04855",
    archivePrefix = "arXiv",
    primaryClass = "gr-qc",
    reportNumber = "LIGO-P151226",
    doi = "10.1103/PhysRevLett.116.241103",
    journal = "Phys. Rev. Lett.",
    volume = "116",
    number = "24",
    pages = "241103",
    year = "2016"
}

@article{EventHorizonTelescope:2022xqj,
    author = "Akiyama, Kazunori and others",
    collaboration = "Event Horizon Telescope",
    title = "{First Sagittarius A* Event Horizon Telescope Results. VI. Testing the Black Hole Metric}",
    eprint = "2311.09484",
    archivePrefix = "arXiv",
    primaryClass = "astro-ph.HE",
    reportNumber = "FERMILAB-PUB-22-422-PPD",
    doi = "10.3847/2041-8213/ac6756",
    journal = "Astrophys. J. Lett.",
    volume = "930",
    number = "2",
    pages = "L17",
    year = "2022"
}

@article{EventHorizonTelescope:2022wkp,
    author = "Akiyama, Kazunori and others",
    collaboration = "Event Horizon Telescope",
    title = "{First Sagittarius A* Event Horizon Telescope Results. I. The Shadow of the Supermassive Black Hole in the Center of the Milky Way}",
    eprint = "2311.08680",
    archivePrefix = "arXiv",
    primaryClass = "astro-ph.HE",
    doi = "10.3847/2041-8213/ac6674",
    journal = "Astrophys. J. Lett.",
    volume = "930",
    number = "2",
    pages = "L12",
    year = "2022"
}

@article{EventHorizonTelescope:2019ths,
    author = "Akiyama, Kazunori and others",
    collaboration = "Event Horizon Telescope",
    title = "{First M87 Event Horizon Telescope Results. IV. Imaging the Central Supermassive Black Hole}",
    eprint = "1906.11241",
    archivePrefix = "arXiv",
    primaryClass = "astro-ph.GA",
    doi = "10.3847/2041-8213/ab0e85",
    journal = "Astrophys. J. Lett.",
    volume = "875",
    number = "1",
    pages = "L4",
    year = "2019"
}

@article{EventHorizonTelescope:2019dse,
    author = "Akiyama, Kazunori and others",
    collaboration = "Event Horizon Telescope",
    title = "{First M87 Event Horizon Telescope Results. I. The Shadow of the Supermassive Black Hole}",
    eprint = "1906.11238",
    archivePrefix = "arXiv",
    primaryClass = "astro-ph.GA",
    doi = "10.3847/2041-8213/ab0ec7",
    journal = "Astrophys. J. Lett.",
    volume = "875",
    pages = "L1",
    year = "2019"
}

@article{Gondolo:1999ef,
    author = "Gondolo, Paolo and Silk, Joseph",
    title = "{Dark matter annihilation at the galactic center}",
    eprint = "astro-ph/9906391",
    archivePrefix = "arXiv",
    reportNumber = "MPI-PHT-99-10, OUAST-99-9",
    doi = "10.1103/PhysRevLett.83.1719",
    journal = "Phys. Rev. Lett.",
    volume = "83",
    pages = "1719--1722",
    year = "1999"
}

@article{Bertone:2005xv,
    author = "Bertone, Gianfranco and Merritt, David",
    title = "{Dark matter dynamics and indirect detection}",
    eprint = "astro-ph/0504422",
    archivePrefix = "arXiv",
    reportNumber = "FERMILAB-PUB-05-051-A",
    doi = "10.1142/S0217732305017391",
    journal = "Mod. Phys. Lett. A",
    volume = "20",
    pages = "1021",
    year = "2005"
}

@article{Salucci:2018hqu,
    author = "Salucci, Paolo",
    title = "{The distribution of dark matter in galaxies}",
    eprint = "1811.08843",
    archivePrefix = "arXiv",
    primaryClass = "astro-ph.GA",
    doi = "10.1007/s00159-018-0113-1",
    journal = "Astron. Astrophys. Rev.",
    volume = "27",
    number = "1",
    pages = "2",
    year = "2019"
}

@article{Bertone:2018krk,
    author = "Bertone, Gianfranco and Tait, M. P., Tim",
    title = "{A new era in the search for dark matter}",
    eprint = "1810.01668",
    archivePrefix = "arXiv",
    primaryClass = "astro-ph.CO",
    doi = "10.1038/s41586-018-0542-z",
    journal = "Nature",
    volume = "562",
    number = "7725",
    pages = "51--56",
    year = "2018"
}

@article{Planck:2018vyg,
    author = "Aghanim, N. and others",
    collaboration = "Planck",
    title = "{Planck 2018 results. VI. Cosmological parameters}",
    eprint = "1807.06209",
    archivePrefix = "arXiv",
    primaryClass = "astro-ph.CO",
    doi = "10.1051/0004-6361/201833910",
    journal = "Astron. Astrophys.",
    volume = "641",
    pages = "A6",
    year = "2020",
    note = "[Erratum: Astron.Astrophys. 652, C4 (2021)]"
}

@article{Rubin:1970zza,
    author = "Rubin, Vera C. and Ford, Jr., W. Kent",
    title = "{Rotation of the Andromeda Nebula from a Spectroscopic Survey of Emission Regions}",
    doi = "10.1086/150317",
    journal = "Astrophys. J.",
    volume = "159",
    pages = "379--403",
    year = "1970"
}

@article{Barausse:2014tra,
    author = "Barausse, Enrico and Cardoso, Vitor and Pani, Paolo",
    title = "{Can environmental effects spoil precision gravitational-wave astrophysics?}",
    eprint = "1404.7149",
    archivePrefix = "arXiv",
    primaryClass = "gr-qc",
    doi = "10.1103/PhysRevD.89.104059",
    journal = "Phys. Rev. D",
    volume = "89",
    number = "10",
    pages = "104059",
    year = "2014"
}

@article{Cardoso:2019rou,
    author = "Cardoso, Vitor and Maselli, Andrea",
    title = "{Constraints on the astrophysical environment of binaries with gravitational-wave observations}",
    eprint = "1909.05870",
    archivePrefix = "arXiv",
    primaryClass = "astro-ph.HE",
    doi = "10.1051/0004-6361/202037654",
    journal = "Astron. Astrophys.",
    volume = "644",
    pages = "A147",
    year = "2020"
}

@article{Zwick:2022dih,
    author = "Zwick, Lorenz and Capelo, Pedro R. and Mayer, Lucio",
    title = "{Priorities in gravitational waveforms for future space-borne detectors: vacuum accuracy or environment?}",
    eprint = "2209.04060",
    archivePrefix = "arXiv",
    primaryClass = "gr-qc",
    doi = "10.1093/mnras/stad707",
    journal = "Mon. Not. Roy. Astron. Soc.",
    volume = "521",
    number = "3",
    pages = "4645--4651",
    year = "2023"
}

@article{Cardoso:2021wlq,
    author = "Cardoso, Vitor and Destounis, Kyriakos and Duque, Francisco and Macedo, Rodrigo Panosso and Maselli, Andrea",
    title = "{Black holes in galaxies: Environmental impact on gravitational-wave generation and propagation}",
    eprint = "2109.00005",
    archivePrefix = "arXiv",
    primaryClass = "gr-qc",
    doi = "10.1103/PhysRevD.105.L061501",
    journal = "Phys. Rev. D",
    volume = "105",
    number = "6",
    pages = "L061501",
    year = "2022"
}

@article{Speeney:2024mas,
    author = "Speeney, Nicholas and Berti, Emanuele and Cardoso, Vitor and Maselli, Andrea",
    title = "{Black holes surrounded by generic matter distributions: Polar perturbations and energy flux}",
    eprint = "2401.00932",
    archivePrefix = "arXiv",
    primaryClass = "gr-qc",
    doi = "10.1103/PhysRevD.109.084068",
    journal = "Phys. Rev. D",
    volume = "109",
    number = "8",
    pages = "084068",
    year = "2024"
}

@article{Maeda:2024tsg,
    author = "Maeda, Kei-ichi and Cardoso, Vitor and Wang, Anzhong",
    title = "{Einstein cluster as central spiky distribution of galactic dark matter}",
    eprint = "2410.04175",
    archivePrefix = "arXiv",
    primaryClass = "gr-qc",
    doi = "10.1103/PhysRevD.111.044060",
    journal = "Phys. Rev. D",
    volume = "111",
    number = "4",
    pages = "044060",
    year = "2025"
}

@article{Shen:2023erj,
    author = "Shen, Zibo and Wang, Anzhong and Gong, Yungui and Yin, Shaoyu",
    title = "{Analytical models of supermassive black holes in galaxies surrounded by dark matter halos}",
    eprint = "2311.12259",
    archivePrefix = "arXiv",
    primaryClass = "gr-qc",
    doi = "10.1016/j.physletb.2024.138797",
    journal = "Phys. Lett. B",
    volume = "855",
    pages = "138797",
    year = "2024"
}

@article{Zhao:2026yis,
    author = "Zhao, Yang and Gong, Yungui",
    title = "{Dark matter distributions around extreme mass ratio inspirals: effects of radial pressure and relativistic treatment}",
    eprint = "2602.12022",
    archivePrefix = "arXiv",
    primaryClass = "gr-qc",
    month = "2",
    year = "2026"
}

@article{Zare:2026lgt,
    author = "Zare, S. and Hosseinifar, F. and Nieto, L. M. and Gogoi, D. J. and Boshkayev, K. and Urazalina, A. and Hassanabadi, H.",
    title = "{Schwarzschild black hole in King{\textquoteright}s dark matter halo}",
    doi = "10.1140/epjc/s10052-025-15261-z",
    journal = "Eur. Phys. J. C",
    volume = "86",
    number = "2",
    pages = "160",
    year = "2026"
}

@article{Boltaev:2026prm,
    author = "Boltaev, Azalbek and Xamidov, Tursunali and Shaymatov, Sanjar",
    title = "{Observational Signatures of Exact Black Hole Solutions in a Dark Matter Halo}",
    eprint = "2603.17986",
    archivePrefix = "arXiv",
    primaryClass = "gr-qc",
    month = "3",
    year = "2026"
}

@article{Konoplya:2022hbl,
    author = "Konoplya, R. A. and Zhidenko, A.",
    title = "{Solutions of the Einstein Equations for a Black Hole Surrounded by a Galactic Halo}",
    eprint = "2202.02205",
    archivePrefix = "arXiv",
    primaryClass = "gr-qc",
    doi = "10.3847/1538-4357/ac76bc",
    journal = "Astrophys. J.",
    volume = "933",
    number = "2",
    pages = "166",
    year = "2022"
}

@article{Ponglertsakul:2026upb,
    author = "Ponglertsakul, Supakchai and Senjaya, David",
    title = "{A Novel Kerr-like Black Hole in a General Double Power Law Dark Matter Environment: Geometry, Spectroscopy, and Energy Extraction}",
    eprint = "2608.05861",
    archivePrefix = "arXiv",
    primaryClass = "gr-qc",
    reportNumber = "7872921E",
    month = "8",
    year = "2026"
}

@article{Fier:2026yjc,
    author = "Fier, Jared and Abdelshahed, Farah and Kowalczyk, Lilly and Lyu, Zhengfei and Wang, Anzhong",
    title = "{Probing the Distribution and Nature of Dark Matter Around Supermassive Black Holes from EMRI and IMRI Gravitational Waves}",
    eprint = "2608.12540",
    archivePrefix = "arXiv",
    primaryClass = "gr-qc",
    month = "8",
    year = "2026"
}

@article{Shen:2024qbb,
    author = "Shen, Zibo and Wang, Anzhong and Yin, Shaoyu",
    title = "{Inner radius and energy conditions of dark matter halos surrounding Schwarzschild black holes}",
    eprint = "2408.05417",
    archivePrefix = "arXiv",
    primaryClass = "gr-qc",
    doi = "10.1016/j.physletb.2025.139300",
    journal = "Phys. Lett. B",
    volume = "862",
    pages = "139300",
    year = "2025"
}

@article{Fernandes:2025osu,
    author = "Fernandes, Pedro G. S. and Cardoso, Vitor",
    title = "{Spinning Black Holes in Astrophysical Environments}",
    eprint = "2507.04389",
    archivePrefix = "arXiv",
    primaryClass = "gr-qc",
    doi = "10.1103/9shv-5d21",
    journal = "Phys. Rev. Lett.",
    volume = "135",
    number = "21",
    pages = "211403",
    year = "2025"
}

@article{Zhang:2024hjr,
    author = "Zhang, Chao and Fu, Guoyang and Zhang, Chunyu",
    title = "{Rotating galactic black holes}",
    eprint = "2403.19933",
    archivePrefix = "arXiv",
    primaryClass = "gr-qc",
    month = "3",
    year = "2024"
}

@article{Fu:2026gcu,
    author = "Fu, Guoyang and Liu, Yunqi and Wu, Jian-Pin and Wang, Bin and Yue, Rui-Hong",
    title = "{Probing Dark Matter with Gravitational Waves: Spin-Modulated Dephasing from Black Holes in Halos}",
    eprint = "2608.15099",
    archivePrefix = "arXiv",
    primaryClass = "gr-qc",
    month = "8",
    year = "2026"
}

@article{Yue:2026gju,
    author = "Yue, Rui-Hong and Zhao, Yu-Qian and Qian, Wei-Liang",
    title = "{A metric solution for rotating black holes embedded in dark matter halos with central spikes}",
    eprint = "2605.15330",
    archivePrefix = "arXiv",
    primaryClass = "gr-qc",
    month = "5",
    year = "2026"
}

@article{Konoplya:2011qq,
    author = "Konoplya, R. A. and Zhidenko, A.",
    title = "{Quasinormal modes of black holes: From astrophysics to string theory}",
    eprint = "1102.4014",
    archivePrefix = "arXiv",
    primaryClass = "gr-qc",
    doi = "10.1103/RevModPhys.83.793",
    journal = "Rev. Mod. Phys.",
    volume = "83",
    pages = "793--836",
    year = "2011"
}

@article{Berti:2009kk,
    author = "Berti, Emanuele and Cardoso, Vitor and Starinets, Andrei O.",
    title = "{Quasinormal modes of black holes and black branes}",
    eprint = "0905.2975",
    archivePrefix = "arXiv",
    primaryClass = "gr-qc",
    doi = "10.1088/0264-9381/26/16/163001",
    journal = "Class. Quant. Grav.",
    volume = "26",
    pages = "163001",
    year = "2009"
}

@article{Kokkotas:1999bd,
    author = "Kokkotas, Kostas D. and Schmidt, Bernd G.",
    title = "{Quasinormal modes of stars and black holes}",
    eprint = "gr-qc/9909058",
    archivePrefix = "arXiv",
    doi = "10.12942/lrr-1999-2",
    journal = "Living Rev. Rel.",
    volume = "2",
    pages = "2",
    year = "1999"
}

@article{Regge:1957td,
    author = "Regge, Tullio and Wheeler, John A.",
    title = "{Stability of a Schwarzschild singularity}",
    doi = "10.1103/PhysRev.108.1063",
    journal = "Phys. Rev.",
    volume = "108",
    pages = "1063--1069",
    year = "1957"
}

@article{Zerilli:1970se,
    author = "Zerilli, Frank J.",
    title = "{Effective potential for even parity Regge-Wheeler gravitational perturbation equations}",
    doi = "10.1103/PhysRevLett.24.737",
    journal = "Phys. Rev. Lett.",
    volume = "24",
    pages = "737--738",
    year = "1970"
}

@article{Kobayashi:2012kh,
    author = "Kobayashi, Tsutomu and Motohashi, Hayato and Suyama, Teruaki",
    title = "{Black hole perturbation in the most general scalar-tensor theory with second-order field equations I: the odd-parity sector}",
    eprint = "1202.4893",
    archivePrefix = "arXiv",
    primaryClass = "gr-qc",
    reportNumber = "RESCEU-3-12, KUNS-2385",
    doi = "10.1103/PhysRevD.85.084025",
    journal = "Phys. Rev. D",
    volume = "85",
    pages = "084025",
    year = "2012",
    note = "[Erratum: Phys.Rev.D 96, 109903 (2017)]"
}

@article{Kobayashi:2014wsa,
    author = "Kobayashi, Tsutomu and Motohashi, Hayato and Suyama, Teruaki",
    title = "{Black hole perturbation in the most general scalar-tensor theory with second-order field equations II: the even-parity sector}",
    eprint = "1402.6740",
    archivePrefix = "arXiv",
    primaryClass = "gr-qc",
    reportNumber = "RESCEU-5-14, RUP-14-4, RESCEU-5/14",
    doi = "10.1103/PhysRevD.89.084042",
    journal = "Phys. Rev. D",
    volume = "89",
    number = "8",
    pages = "084042",
    year = "2014"
}

@article{Ganguly:2017ort,
    author = "Ganguly, Apratim and Gannouji, Radouane and Gonzalez-Espinoza, Manuel and Pizarro-Moya, Carlos",
    title = "{Black hole stability under odd-parity perturbations in Horndeski gravity}",
    eprint = "1710.07669",
    archivePrefix = "arXiv",
    primaryClass = "gr-qc",
    doi = "10.1088/1361-6382/aac8a0",
    journal = "Class. Quant. Grav.",
    volume = "35",
    number = "14",
    pages = "145008",
    year = "2018"
}

@article{Antoniou:2024hlf,
    author = "Antoniou, Georgios",
    title = "{Quasinormal modes of hairy black holes in shift-symmetric theories}",
    eprint = "2406.01508",
    archivePrefix = "arXiv",
    primaryClass = "gr-qc",
    doi = "10.1103/PhysRevD.110.044029",
    journal = "Phys. Rev. D",
    volume = "110",
    number = "4",
    pages = "044029",
    year = "2024"
}

@article{Antoniou:2024jku,
    author = "Antoniou, Georgios and Gualtieri, Leonardo and Pani, Paolo",
    title = "{Gravitational quasinormal modes of black holes in quadratic gravity}",
    eprint = "2412.15037",
    archivePrefix = "arXiv",
    primaryClass = "gr-qc",
    doi = "10.1103/PhysRevD.111.064059",
    journal = "Phys. Rev. D",
    volume = "111",
    number = "6",
    pages = "064059",
    year = "2025"
}

@article{Antoniou:2024gdf,
    author = "Antoniou, Georgios and Macedo, Caio F. B. and Maselli, Andrea and Sotiriou, Thomas P.",
    title = "{Axial perturbations of hairy black holes in generalized scalar-tensor theories}",
    eprint = "2404.02479",
    archivePrefix = "arXiv",
    primaryClass = "gr-qc",
    doi = "10.1103/PhysRevD.110.024065",
    journal = "Phys. Rev. D",
    volume = "110",
    number = "2",
    pages = "024065",
    year = "2024"
}

@article{Antoniou:2026vrm,
    author = "Antoniou, Georgios",
    title = "{Perturbations of black holes with primary hair: time evolutions, quasinormal modes and greybody factors}",
    eprint = "2608.26984",
    archivePrefix = "arXiv",
    primaryClass = "gr-qc",
    month = "8",
    year = "2026"
}

@article{Blazquez-Salcedo:2016enn,
    author = "Bl{\'a}zquez-Salcedo, Jose Luis and Macedo, Caio F. B. and Cardoso, Vitor and Ferrari, Valeria and Gualtieri, Leonardo and Khoo, Fech Scen and Kunz, Jutta and Pani, Paolo",
    title = "{Perturbed black holes in Einstein-dilaton-Gauss-Bonnet gravity: Stability, ringdown, and gravitational-wave emission}",
    eprint = "1609.01286",
    archivePrefix = "arXiv",
    primaryClass = "gr-qc",
    doi = "10.1103/PhysRevD.94.104024",
    journal = "Phys. Rev. D",
    volume = "94",
    number = "10",
    pages = "104024",
    year = "2016"
}

@article{Molina:2010fb,
    author = "Molina, C. and Pani, Paolo and Cardoso, Vitor and Gualtieri, Leonardo",
    title = "{Gravitational signature of Schwarzschild black holes in dynamical Chern-Simons gravity}",
    eprint = "1004.4007",
    archivePrefix = "arXiv",
    primaryClass = "gr-qc",
    doi = "10.1103/PhysRevD.81.124021",
    journal = "Phys. Rev. D",
    volume = "81",
    pages = "124021",
    year = "2010"
}

@article{Konoplya:2026cus,
    author = "Konoplya, Roman and Stashko, Oleksandr and Stuchl{\'\i}k, Zden{\v{e}}k",
    title = "{Quasinormal Modes and Hawking Radiation of Black Holes with Primary Scalar Hair}",
    eprint = "2606.16125",
    archivePrefix = "arXiv",
    primaryClass = "gr-qc",
    month = "6",
    year = "2026"
}

@article{Xiong:2023usm,
    author = "Xiong, Wei and Li, Peng-Cheng",
    title = "{Quasinormal modes of rotating accelerating black holes}",
    eprint = "2305.04040",
    archivePrefix = "arXiv",
    primaryClass = "gr-qc",
    doi = "10.1103/PhysRevD.108.044064",
    journal = "Phys. Rev. D",
    volume = "108",
    number = "4",
    pages = "044064",
    year = "2023"
}

@article{Konoplya:2023hqb,
    author = "Konoplya, R. A.",
    title = "{The sound of the event horizon}",
    eprint = "2312.16249",
    archivePrefix = "arXiv",
    primaryClass = "gr-qc",
    doi = "10.1142/S0218271823420142",
    journal = "Int. J. Mod. Phys. D",
    volume = "32",
    number = "14",
    pages = "2342014",
    year = "2023"
}

@article{Konoplya:2023ahd,
    author = "Konoplya, R. A. and Ovchinnikov, D. and Ahmedov, B.",
    title = "{Bardeen spacetime as a quantum corrected Schwarzschild black hole: Quasinormal modes and Hawking radiation}",
    eprint = "2307.10801",
    archivePrefix = "arXiv",
    primaryClass = "gr-qc",
    doi = "10.1103/PhysRevD.108.104054",
    journal = "Phys. Rev. D",
    volume = "108",
    number = "10",
    pages = "104054",
    year = "2023"
}

@article{Konoplya:2022hll,
    author = "Konoplya, R. A. and Zinhailo, A. F. and Kunz, J. and Stuchlik, Z. and Zhidenko, A.",
    title = "{Quasinormal ringing of regular black holes in asymptotically safe gravity: the importance of overtones}",
    eprint = "2206.14714",
    archivePrefix = "arXiv",
    primaryClass = "gr-qc",
    doi = "10.1088/1475-7516/2022/10/091",
    journal = "JCAP",
    volume = "10",
    pages = "091",
    year = "2022"
}

@article{Konoplya:2022pbc,
    author = "Konoplya, R. A. and Zhidenko, A.",
    title = "{First few overtones probe the event horizon geometry}",
    eprint = "2209.00679",
    archivePrefix = "arXiv",
    primaryClass = "gr-qc",
    doi = "10.1016/j.jheap.2024.10.015",
    journal = "JHEAp",
    volume = "44",
    pages = "419--426",
    year = "2024"
}

@article{Fu:2023drp,
    author = "Fu, Guoyang and Zhang, Dan and Liu, Peng and Kuang, Xiao-Mei and Wu, Jian-Pin",
    title = "{Peculiar properties in quasinormal spectra from loop quantum gravity effect}",
    eprint = "2301.08421",
    archivePrefix = "arXiv",
    primaryClass = "gr-qc",
    doi = "10.1103/PhysRevD.109.026010",
    journal = "Phys. Rev. D",
    volume = "109",
    number = "2",
    pages = "026010",
    year = "2024"
}

@article{Gong:2023ghh,
    author = "Gong, Huajie and Li, Shulan and Zhang, Dan and Fu, Guoyang and Wu, Jian-Pin",
    title = "{Quasinormal modes of quantum-corrected black holes}",
    eprint = "2312.17639",
    archivePrefix = "arXiv",
    primaryClass = "gr-qc",
    doi = "10.1103/PhysRevD.110.044040",
    journal = "Phys. Rev. D",
    volume = "110",
    number = "4",
    pages = "044040",
    year = "2024"
}

@article{Zhang:2024nny,
    author = "Zhang, Dan and Gong, Huajie and Fu, Guoyang and Wu, Jian-Pin and Pan, Qiyuan",
    title = "{Quasinormal modes of a regular black hole with sub-Planckian curvature}",
    eprint = "2402.15085",
    archivePrefix = "arXiv",
    primaryClass = "gr-qc",
    doi = "10.1140/epjc/s10052-024-12928-x",
    journal = "Eur. Phys. J. C",
    volume = "84",
    number = "6",
    pages = "564",
    year = "2024"
}

@article{Zhu:2024wic,
    author = "Zhu, Li-Gang and Fu, Guoyang and Li, Shulan and Zhang, Dan and Wu, Jian-Pin",
    title = "{Quasinormal modes of a charged loop quantum black hole}",
    eprint = "2410.00543",
    archivePrefix = "arXiv",
    primaryClass = "gr-qc",
    doi = "10.1103/PhysRevD.111.104008",
    journal = "Phys. Rev. D",
    volume = "111",
    number = "10",
    pages = "104008",
    year = "2025"
}

@article{Dong:2024ams,
    author = "Dong, Zhongzhinan and Zhang, Dan and Fu, Guoyang and Wu, Jian-Pin",
    title = "{Quasinormal modes of a d-dimensional regular black hole featuring an integrable singularity}",
    eprint = "2412.20457",
    archivePrefix = "arXiv",
    primaryClass = "gr-qc",
    doi = "10.1140/epjc/s10052-025-13926-3",
    journal = "Eur. Phys. J. C",
    volume = "85",
    number = "2",
    pages = "215",
    year = "2025"
}

@article{Zhang:2025ygb,
    author = "Zhang, Dan and Tan, Qin and Fu, Guoyang and Gong, Huajie and Wu, Jian-Pin and Pan, Qiyuan",
    title = "{Echoes from the Minkowski-core spacetime}",
    eprint = "2509.23215",
    archivePrefix = "arXiv",
    primaryClass = "gr-qc",
    doi = "10.1007/s11433-025-2904-2",
    journal = "Sci. China Phys. Mech. Astron.",
    volume = "69",
    number = "5",
    pages = "250412",
    year = "2026"
}

@article{Tang:2024txx,
    author = "Tang, Chen and Ling, Yi and Jiang, Qing-Quan and Li, Guo-Ping",
    title = "{Quasinormal modes of regular black holes with sub-Planckian curvature and Minkowskian core}",
    eprint = "2411.01764",
    archivePrefix = "arXiv",
    primaryClass = "gr-qc",
    doi = "10.1140/epjc/s10052-024-13699-1",
    journal = "Eur. Phys. J. C",
    volume = "84",
    number = "12",
    pages = "1296",
    year = "2024"
}

@article{Pezzella:2024tkf,
    author = "Pezzella, Laura and Destounis, Kyriakos and Maselli, Andrea and Cardoso, Vitor",
    title = "{Quasinormal modes of black holes embedded in halos of matter}",
    eprint = "2412.18651",
    archivePrefix = "arXiv",
    primaryClass = "gr-qc",
    doi = "10.1103/PhysRevD.111.064026",
    journal = "Phys. Rev. D",
    volume = "111",
    number = "6",
    pages = "064026",
    year = "2025"
}

@article{Konoplya:2021ube,
    author = "Konoplya, R. A.",
    title = "{Black holes in galactic centers: Quasinormal ringing, grey-body factors and Unruh temperature}",
    eprint = "2109.01640",
    archivePrefix = "arXiv",
    primaryClass = "gr-qc",
    doi = "10.1016/j.physletb.2021.136734",
    journal = "Phys. Lett. B",
    volume = "823",
    pages = "136734",
    year = "2021"
}

@article{Feng:2025iao,
    author = "Feng, Xing-Hui and Zhang, Guang-Yu",
    title = "{Shadow and quasi-normal modes of Schwarzschild{\textendash}Hernquist black hole}",
    eprint = "2509.04001",
    archivePrefix = "arXiv",
    primaryClass = "gr-qc",
    doi = "10.1140/epjc/s10052-026-15293-z",
    journal = "Eur. Phys. J. C",
    volume = "86",
    number = "1",
    pages = "36",
    year = "2026"
}

@article{Zhao:2026eti,
    author = "Zhao, Yu-Qian and Pani, Paolo",
    title = "{Quasinormal modes and tidal responses of black holes in generic anisotropic matter environments}",
    eprint = "2606.11380",
    archivePrefix = "arXiv",
    primaryClass = "gr-qc",
    month = "6",
    year = "2026"
}

@article{Liu:2024xcd,
    author = "Liu, Dong and Yang, Yi and Long, Zheng-Wen",
    title = "{Probing black holes in a dark matter spike of M87 using quasinormal modes}",
    eprint = "2401.09182",
    archivePrefix = "arXiv",
    primaryClass = "gr-qc",
    doi = "10.1140/epjc/s10052-024-13096-8",
    journal = "Eur. Phys. J. C",
    volume = "84",
    number = "7",
    pages = "731",
    year = "2024"
}

@article{Zhang:2021bdr,
    author = "Zhang, Chao and Zhu, Tao and Wang, Anzhong",
    title = "{Gravitational axial perturbations of Schwarzschild-like black holes in dark matter halos}",
    eprint = "2111.04966",
    archivePrefix = "arXiv",
    primaryClass = "gr-qc",
    doi = "10.1103/PhysRevD.104.124082",
    journal = "Phys. Rev. D",
    volume = "104",
    number = "12",
    pages = "124082",
    year = "2021"
}

@article{Nollert:1996rf,
    author = "Nollert, Hans-Peter",
    title = "{About the significance of quasinormal modes of black holes}",
    eprint = "gr-qc/9602032",
    archivePrefix = "arXiv",
    doi = "10.1103/PhysRevD.53.4397",
    journal = "Phys. Rev. D",
    volume = "53",
    pages = "4397--4402",
    year = "1996"
}

@article{Nollert:1998ys,
    author = "Nollert, Hans-Peter and Price, Richard H.",
    title = "{Quantifying excitations of quasinormal mode systems}",
    eprint = "gr-qc/9810074",
    archivePrefix = "arXiv",
    doi = "10.1063/1.532698",
    journal = "J. Math. Phys.",
    volume = "40",
    pages = "980--1010",
    year = "1999"
}

@article{Jaramillo:2020tuu,
    author = "Jaramillo, Jos{\'e} Luis and Panosso Macedo, Rodrigo and Al Sheikh, Lamis",
    title = "{Pseudospectrum and Black Hole Quasinormal Mode Instability}",
    eprint = "2004.06434",
    archivePrefix = "arXiv",
    primaryClass = "gr-qc",
    doi = "10.1103/PhysRevX.11.031003",
    journal = "Phys. Rev. X",
    volume = "11",
    number = "3",
    pages = "031003",
    year = "2021"
}

@article{Jaramillo:2021tmt,
    author = "Jaramillo, Jos{\'e} Luis and Panosso Macedo, Rodrigo and Sheikh, Lamis Al",
    title = "{Gravitational Wave Signatures of Black Hole Quasinormal Mode Instability}",
    eprint = "2105.03451",
    archivePrefix = "arXiv",
    primaryClass = "gr-qc",
    doi = "10.1103/PhysRevLett.128.211102",
    journal = "Phys. Rev. Lett.",
    volume = "128",
    number = "21",
    pages = "211102",
    year = "2022"
}

@article{Cheung:2021bol,
    author = "Cheung, Mark Ho-Yeuk and Destounis, Kyriakos and Macedo, Rodrigo Panosso and Berti, Emanuele and Cardoso, Vitor",
    title = "{Destabilizing the Fundamental Mode of Black Holes: The Elephant and the Flea}",
    eprint = "2111.05415",
    archivePrefix = "arXiv",
    primaryClass = "gr-qc",
    doi = "10.1103/PhysRevLett.128.111103",
    journal = "Phys. Rev. Lett.",
    volume = "128",
    number = "11",
    pages = "111103",
    year = "2022"
}

@article{Berti:2022xfj,
    author = "Berti, Emanuele and Cardoso, Vitor and Cheung, Mark Ho-Yeuk and Di Filippo, Francesco and Duque, Francisco and Martens, Paul and Mukohyama, Shinji",
    title = "{Stability of the fundamental quasinormal mode in time-domain observations against small perturbations}",
    eprint = "2205.08547",
    archivePrefix = "arXiv",
    primaryClass = "gr-qc",
    doi = "10.1103/PhysRevD.106.084011",
    journal = "Phys. Rev. D",
    volume = "106",
    number = "8",
    pages = "084011",
    year = "2022"
}

@book{Bosch_White_2010,
    author = "Mo, Houjun and van den Bosch, Frank and White, Simon",
    title = "{Galaxy Formation and Evolution}",
    address = "Cambridge",
    publisher = "Cambridge University Press",
    year = "2010"
}
\end{document}